\pdfoutput=1
\documentclass[nonacm,sigconf]{acmart}

\usepackage{algorithm}
\usepackage[noend]{algpseudocode}
\algnewcommand{\algorithmicassumption}{\textbf{Requirement:}}
\algnewcommand{\Assume}{\item[\algorithmicassumption]}
\algrenewcomment[1]{\hfill \(\triangleright\) \emph{\small #1}} 
\algnewcommand{\InlineIf}[2]{
  \algorithmicif\ #1\ \algorithmicthen\ #2}
\algnewcommand{\InlineElse}[1]{
  \algorithmicelse\ #1}
\algnewcommand{\InlineIfElse}[3]{
  \algorithmicif\ #1\ \algorithmicthen\ #2\ \algorithmicelse\ #3}
\algnewcommand{\InlineFor}[2]{\algorithmicfor\ #1\ \algorithmicdo\ #2} 
\algnewcommand{\CommentLine}[1]{\(\triangleright\) \emph{\small #1}}

\algnewcommand{\algorithmicand}{\textbf{and}}
\algnewcommand{\algorithmicor}{\textbf{or}}
\algnewcommand{\FOR}{\algorithmicfor}
\algnewcommand{\OR}{\algorithmicor}
\algnewcommand{\AND}{\algorithmicand}
\algnewcommand{\IF}{\algorithmicif}
\algnewcommand{\THEN}{\algorithmicthen}
\algnewcommand{\ELSE}{\algorithmicelse}

\usepackage[shortlabels]{enumitem}
\usepackage[capitalise]{cleveref}
\usepackage{mathtools}
\usepackage{graphicx}
\graphicspath{{data}}
\usepackage{xcolor}

\makeatletter
\newsavebox{\@brx}
\newcommand{\llangle}[1][]{\savebox{\@brx}{\(\m@th{#1\langle}\)}%
  \mathopen{\copy\@brx\kern-0.5\wd\@brx\usebox{\@brx}}}
\newcommand{\rrangle}[1][]{\savebox{\@brx}{\(\m@th{#1\rangle}\)}%
  \mathclose{\copy\@brx\kern-0.5\wd\@brx\usebox{\@brx}}}

\newcommand{\ring}{\mathcal{R}}
\newcommand{\ideal}{\mathcal{I}}
\newcommand{\idealJ}{\mathcal{J}}
\newcommand{\module}{\mathcal{M}}
\newcommand{\genby}[1]{\left\langle #1 \right\rangle}
\newcommand{\modgenby}[1]{\llangle #1 \rrangle}
\newcommand{\modf}{f}
\newcommand{\modF}{F}
\newcommand{\tensoralgebra}[1][\module]{T(#1)}
\NewDocumentCommand\extalg{O{}O{\module}}{\bigwedge\nolimits^{#1}(#2)}
\newcommand{\dual}[1]{\left({#1}\right)^*}
\newcommand{\field}{\Bbbk}
\newcommand{\algclosure}{\bar{\Bbbk}}
\newcommand{\ZZ}{\mathbb{Z}}
\newcommand{\ZZpos}{\ZZ_{>0}}
\newcommand{\ZZnonneg}{\ZZ_{\ge 0}}
\newcommand{\nvars}{n}
\newcommand{\pring}{\field[x_1,\dots,x_\nvars]}
\newcommand{\pringshort}[1][\nvars]{\ring_{#1}}
\NewDocumentCommand\Mon{O{d}O{\pringshort}}{\operatorname{Mon}_{#1}\left(#2\right)}
\newcommand{\ord}{\succ}
\newcommand{\ordpot}{\ord^{\text{POT}}}
\newcommand{\lexgreater}{\ord_{\text{lex}}}

\newcommand{\Tor}{\operatorname{Tor}}
\DeclareMathOperator{\Hom}{Hom}

\newcommand{\affspace}[1]{\mathbb{A}^{#1}}
\newcommand{\genericf}[3]{\mathfrak{f}^{#1}_{(#2,#3)}}
\newcommand{\frakc}{\mathfrak{c}}
\newcommand{\specialseq}[3]{\phi_{#1}(\genericf{}{#2}{#3})}

\newcommand{\specialmat}[3]{\phi_{#1}(\genericf{}{#2}{#3})}

\newcommand{\jac}{\operatorname{jac}}
\newcommand{\critideal}{\mathcal{I}(g,F)}

\newcommand{\nrows}{p}
\newcommand{\ncols}{q}
\newcommand{\polymat}{A} 
\newcommand{\polymatentry}{a} 
\newcommand{\detideal}[2]{I_{#1}\left(#2\right)} 
\newcommand{\detsystem}[2]{F_{#1}\left(#2\right)} 
\newcommand{\maxminorsideal}[2][\nrows]{\detideal{#1}{#2}} 
\newcommand{\maxminorssystem}[2][\nrows]{\detsystem{#1}{#2}} 
\newcommand{\detidealSmallPar}[2]{I_{#1}(#2)} 
 
\newcommand{\maxminorsidealSmallPar}[2][\nrows]{\detidealSmallPar{#1}{#2}}

\DeclareMathOperator{\EN}{\mathbf{EN}}
\DeclareMathOperator{\koszul}{\mathcal{K}}
\DeclareMathOperator{\grade}{grade}
\DeclareMathOperator{\Sym}{Sym}
\DeclareMathOperator{\rank}{rank}
\DeclareMathOperator{\im}{im}
\newcommand{\freeresmod}{\mathcal{E}}
\DeclareMathOperator{\Syz}{Syz}
\DeclareMathOperator{\HF}{HF}

\DeclareMathOperator{\LM}{LM_{\succ}}
\DeclareMathOperator{\LMmod}{LM_{\succ}^{POT}}
\newcommand{\macmat}{\mathscr{M}}
\newcommand{\macmatred}{\widetilde{\mathscr{M}}}
\newcommand{\rows}{\operatorname{rows}}

\DeclareMathOperator{\SigGB}{MatrixF_5}
\DeclareMathOperator{\MaxMinorsSigGB}{MaxDetMatrixF_5}
\DeclareMathOperator{\CritGB}{CritGB}

\newcommand{\myparagraph}[1]{\smallskip\emph{#1.}} 
\newcommand{\numrows}{\mathcal{R}}

\author{Sriram Gopalakrishnan}
\affiliation{%
	\institution{Sorbonne  Universit\'e, \textsc{CNRS}, \textsc{LIP6}}
	\city{F-75005 Paris}
	\postcode{75252}\country{France}}
\additionalaffiliation{%
	\institution{University of Waterloo}
	\city{Waterloo, ON, Canada}
}

\author{Vincent Neiger}
\affiliation{%
	\institution{Sorbonne Universit\'e, \textsc{CNRS}, \textsc{LIP6}}
	\city{F-75005 Paris}
	\postcode{75252}\country{France}}

\author{Mohab Safey El Din}
\affiliation{%
	\institution{Sorbonne Universit\'e, \textsc{CNRS},  \textsc{LIP6}}
	\city{F-75005 Paris}
	\postcode{75252}\country{France}}

\renewcommand{\shortauthors}{S. Gopalakrishnan, V. Neiger, and M. Safey El Din}
\title{Optimized Gröbner basis algorithms for maximal determinantal ideals and critical point computations}

\renewcommand\footnotetextcopyrightpermission[1]{} 

\begin{document}

\theoremstyle{remark}
\newtheorem{remark}[theorem]{Remark}

\begin{abstract}
	Given polynomials $g$ and $f_1,\dots,f_p$, all in
	$\Bbbk[x_1,\dots,x_n]$ for some field $\Bbbk$, we consider the problem
	of computing the critical points of the restriction of $g$ to the
	variety defined by $f_1=\cdots=f_p=0$. These are defined by the
	simultaneous vanishing of the $f_i$'s and all maximal minors of the
	Jacobian matrix associated to $(g,f_1, \ldots, f_p)$. We use the
	Eagon-Northcott complex associated to the ideal generated by these
	maximal minors to gain insight into the syzygy module of the system
	defining these critical points. We devise new $F_5$-type criteria to
	predict and avoid more reductions to zero when computing a Gr\"obner
	basis for the defining system of this critical locus. We give a bound
	for the arithmetic complexity of this enhanced $F_5$ algorithm and
	compare it to the best previously known bound for computing
	critical points using Gr\"obner bases.
\end{abstract}

\maketitle

\section{Introduction}
\label{sec:intro}

\myparagraph{Motivation and problem} Let $\nvars\in\ZZpos$, $\field$ be a field
with algebraic closure $\algclosure$, and $\pringshort = \pring$ be the ring of
polynomials in $x_1, \ldots, x_n$ with coefficients in $\field$.
Consider a sequence $F=(f_1,\dots,f_\nrows)$ of polynomials and another
polynomial $g$, all of them in $\pringshort$, and the Jacobian matrix $\jac(g,F)$
associated to $g$ and $F$. We denote by $\genby{F}$ the ideal of $\pringshort$
generated by $F$, and by $\maxminorsideal[\nrows+1]{\jac(g,F)}$ the ideal
generated by the maximal minors of $\jac(g,F)$.
We consider the problem of computing a \emph{Gröbner basis} of the
ideal
\[
  \critideal=\genby{F}+\maxminorsideal[\nrows+1]{\jac(g,F)}.
\]
When $\genby{F}$ is radical, is equidimensional of codimension $\nrows$, and
defines a smooth algebraic set $V(F)$ in $\algclosure^{\nvars}$, the algebraic
set $V(\critideal)$ in $\algclosure^{\nvars}$ defined by $\critideal$ is the
set of critical points of the restriction of the polynomial map defined by $g$
to $V(F)$.
Such sets arise in many areas such as polynomial optimization \cite{GreSa11,
  GreSa14}, real algebraic geometry \cite{SaSc03, SaSc17, LeSa21, LeSa22} and
their applications in sciences such as robotics \cite{trutman:hal-02905816,
  capco:hal-02925478, capco:hal-03389500, chablat:hal-03596704} and biology
\cite{kaihnsa:hal-02925505, yabo:hal-03671432}.

\myparagraph{Gr\"obner bases} Throughout the paper, we assume that the set of
critical points under consideration is finite. To compute these critical
points, we solve the system consisting of $(f_1,\dots,f_\nrows)$ and the maximal
minors of $\jac(g,F)$. While several recently developed algorithms for solving such systems
 use symbolic homotopies (see e.g.\ \cite{HSSV21, LSSV21}), we
focus here on algebraic algorithms, based on
Gröbner bases. These are central in the area of polynomial system solving
through computer algebra.
We refer to \cite{CoxLittleOShea2007} for a reference textbook on Gröbner
bases. The classical two-step solving strategy consists in first computing a
Gröbner basis for $\critideal$ with respect to the graded reverse lexicographic
(grevlex) order, and then using a change of order algorithm to obtain a
lexicographic Gr\"obner basis for $\critideal$, from which the solutions can be
read off.

Our focus in this paper is on the first of these two steps, which is nowadays
frequently the most expensive of the two \cite{FaugereMou2017,berthomieu:hal-03580736}.

Evolutions of Buchberger's original Gröbner basis algorithm
\cite{bGroebner1965} have led to linear algebra-based algorithms,
which go back to Lazard's algorithm
\cite{Lazard1983} and include the now standard $F_4$ and $F_5$ algorithms \cite{Faugere1999, Faugere2002}
which have shown their practical efficiency.

These algorithms work by row echelonization of \emph{Macau\-lay
matrices}, whose columns are indexed by the
monomials of $\pringshort$ up to some degree $d$ and sorted by grevlex, and
whose rows store the coefficients of the input polynomials multiplied by the
monomials required to reach the degree $d$. If $d$ is large enough, the obtained
echelon form yields a Gröbner basis \cite{Lazard1983}. This large enough degree is often
called \emph{degree of regularity}. Successive enhancements of this approach have
culminated with the $F_5$ algorithm \cite{Faugere2002} (see
also \cite{EderFaugere2016}), which manages to a priori discard rows that
would otherwise reduce to $0$ upon echelonization. It has been shown that for sequences of
polynomials that are \emph{generic} (in the sense of the Zariski topology),
the so-called $F_5$-criterion detects all reductions to $0$ a priori, and
$F_5$ thus saves all computations related to them.
A key observation behind this criterion is that
these reductions to $0$ come from the Koszul syzygies, induced by the
commutativity of the multiplication in $\pringshort$. This yields faster
Gröbner basis computations for ideals generated by such generic sequences
\cite{BardetFaugereSalvy2015}.

However, it is not the case that the $F_5$-criterion eliminates all reductions
to $0$ on classes of structured systems, including the ones defining critical
points. For these systems, it has been established \cite[Thm.\,3.4]{Spa14}
\cite{FaSaSpa12} that under genericity assumptions on $(g,F)$,
a \emph{grevlex} Gröbner basis of \(\critideal\) can be computed using 
\[
O\left( \left( p+\binom{n}{p+1}
  \right)\binom{n+(n+p)d_0+1}{n}^\omega \right)
\] 
operations in $\field$; this is done by determining the degree of regularity of the
ideal. (Here, $\omega>2$ is a feasible exponent for square matrix
multiplication over \(\field\).) The goal of this paper is to introduce a criterion, for critical point
systems, that complements the $F_5$-criterion so as to avoid more
reductions to \(0\) and thus gain in efficiency.

\myparagraph{Contributions} 
It is known that the $F_5$ algorithm can be enhanced with some insight into the
syzygy modules associated to the generators of the ideal under study
\cite{EderFaugere2016}. This is exploited in \cite[Algo.\,3]{GoNeSa23}, where a
free resolution is used to obtain generators for each syzygy module, allowing
then to call the $F_5$ algorithm to compute Gr\"obner bases without reductions
to \(0\) for the syzygy modules and finally for the ideal itself. The latter
reference studies the case of square matrices with rank deficiency, which leads
to considering free resolutions of a fixed length, whose boundary homomorphisms
admit transparent enough descriptions that computing syzygy modules from them
is a straightforward process. In contrast, here we have to deal with a more
involved complex, namely the Eagon-Northcott one \cite{EagonNorthcott1962},
whose length depends on the size of the matrix under consideration. It is thus
not clear that computing Gr\"obner bases for the syzygy modules could lead to
an efficient algorithm. Still, the specific nature of syzygies between maximal
minors allows us to take a more sophisticated approach for the detection of
reductions to \(0\).

We actually analyze the first syzygy module of the Eagon-North\-cott complex
and exhibit a submodule of its leading terms (w.r.t.\ some module ordering
induced by grevlex). This has a simple algorithmic consequence: by
incrementally computing Gröbner bases of ideals generated by the leftmost
entries of the considered matrix (which fits perfectly with the incremental
nature of $F_5$), one obtains enough information to easily identify a submodule
of the one generated by the leading terms of the first syzygy module. When
combined with the syzygy criterion of $F_5$ (see
\cite[Lemma\,6.4]{EderFaugere2016}), this allows us to discard a significant
number of rows in the Macaulay matrices that arise when computing critical
points. This technique can also be used for pure determinantal ideals, i.e.\
ideals generated maximal minors of a given matrix with entries in
$\pringshort$. Hence, all in all, we obtain a new $F_5$-type algorithm
dedicated to systems involving the maximal minors of a matrix with entries in
$\pringshort$ that avoids some reductions to $0$ that the $F_5$-criterion alone
does not avoid.

Quantifying the resulting complexity gain is challenging. As usual for
analyzing Gröbner basis algorithms, one needs genericity assumptions. Here,
genericity regards the coefficients of $F$ and $g$, and we assume a variant of
Fröberg's conjecture. We show that the extra computations performed to identify
some of the leading terms of the first syzygy module is negligible compared to
the cost of the whole computation. To obtain a complexity estimate, we count
those leading terms, which provides a lower bound on the number of rows of the
Macaulay matrices which our approach discards. The obtained formula is rather
involved, but much more precise than an analysis based on the degree of
regularity alone.

Our complexity analysis does not take into account all rows removed by the full
syzygy criterion. Hence, it is plausible that our complexity bound may be
improved in the future. Since the complexity bound that we give is rather
involved, we evaluate the number of rows in the Macaulay matrices that we build
for certain parameters. Comparing this count to the upper bound on the number
of rows built by Lazard's algorithm obtained in \cite[Theorem\,3.4]{Spa14}, we
see that the complexity bound improvement that we obtain is at least polynomial
in $\nvars$ and that, if we were able to take into account the full syzygy
criterion, it may be exponential in $\nvars$.

\myparagraph{Outline} Basic notions from algebra
and signature Gröbner bases are recalled
in \cref{sec:prelim,sec:grobner}, respectively. In
\cref{sec:eagonnorthcott}, we present constructions on which the
Eagon-Northcott complex relies and show how to use them to obtain a new F5-type
criterion. In \cref{sec:criticalpoints}, we apply this criterion to design a
Gröbner basis algorithm dedicated to critical points. Finally,
\cref{sec:complexity} carries out a complexity analysis of that
algorithm under genericity assumptions.

\section{Preliminaries}
\label{sec:prelim}

In this section, we recall the basic constructions and establish the notation
upon which we rely throughout the paper.

\myparagraph{Polynomials and matrices}
We denote by $\pringshort=\pring$ the ring of polynomials in $\nvars$
indeterminates over $\field$. For a module $\module$ over a ring $\ring$ and a
subset $F\subseteq\module$, we denote by $\modgenby{F}$ the $\ring$-submodule of
$\module$ generated by $F$. In particular, if $\module=\ring$, so that
$F\subseteq\ring$ is a collection of elements of $\ring$, the $\ring$-submodule
$\modgenby{F}$ of $\ring$ is the ideal $\genby{F}$ of $\ring$ generated by $F$.

For $\alpha\in\ZZnonneg^{n}$, we take $x^{\alpha}=x_1^{\alpha_1}\cdots
x_n^{\alpha_n}\in\pringshort$. For $d\in\ZZnonneg$, we denote by $\Mon$ the set
of monomials of $\pringshort$ of degree $d$.

For a ring $\ring$, we will denote by $\ring^{\nrows\times\ncols}$ the
set of matrices with $\nrows$ rows and $\ncols$ columns with entries
in $\ring$; this is a free $\ring$-module of
rank $\nrows\cdot\ncols$. 
Let $A\in\ring^{\nrows\times\ncols}$, and let $r \in \{1, \ldots, \min(p,q)\}$.
Let $1\le i_1<\cdots<i_r\le p$ and $1\le j_1<\cdots<j_r\le q$ be
two strictly increasing sequences of integers. We denote by $[i_1\cdots i_r\mid
j_1\cdots j_r]_\polymat$ the $r\times r$ submatrix of $A$ with rows indexed by
$(i_1,\dots,i_r)$ and columns indexed by $(j_1,\dots, j_r)$. We denote by
$\detsystem{r}{A}$ the subset of $\ring$ consisting of the minors of $A$ of
size $r\times r$, and by $\detideal{r}{A}=\langle\detsystem{r}{A}\rangle$ the
ideal of $\ring$ generated by $\detsystem{r}{A}$.

\myparagraph{Modules and bases}
In order to introduce the portions of the Eagon-Northcott complex which are
relevant to us, we will need to briefly use the language of tensor, symmetric,
and exterior algebras (we refer to \cite[Chap.\,16,19]{Lang02} as a reference
book on these topics). As such, we introduce their notation and canonical
bases.

A ring $\ring$ is called \emph{graded} if, for each integer $d\ge 0$, there
exist additive abelian groups $\ring_{[d]}$ such that
$\ring=\bigoplus_{d=0}^{\infty}\ring_{[d]}$ and
$\ring_{[d]}\ring_{[e]}\subseteq\ring_{[d+e]}$. The elements of $\ring_{[d]}$
are called the \emph{homogeneous elements of degree $d$}. Our prototypical
example of a graded ring will be the ring $\pringshort$. Here, $\pring_{[d]}$
consists of the homogeneous polynomials of degree $d$ (together with $0$, which
is, by definition, homogeneous of every degree).

A module $\module$ over a graded ring $\ring$ is called \emph{graded} if, for
each $d\ge 0$, there exist abelian groups $\module_{[d]}$ such that
$\module=\bigoplus_{d=0}^{\infty}\module_{[d]}$ and
$\ring_{[d]}\module_{[e]}\subseteq\module_{[d+e]}$. For an integer $s\ge 1$,
$\ring^{s}$ naturally carries the structure of a free $\ring$-module of rank
$s$. We take as a basis for $\ring^{s}$ the standard basis vectors $\{e_i:1\le
i\le s\}$. If $\ring$ is graded, it induces a natural grading on all free
modules $\ring^{s}$. For a graded module $\module$ and $e\in\ZZ$, we denote by
$\module(e)$ the module $\module$ with the grading such that
$\module(e)_{[d]}=\module_{[d+e]}$.

For $s\in\ZZpos$, we call an element of $\pringshort^{s}$ a \emph{monomial} if
it takes the form $x^{\alpha}e_i$ for some $\alpha\in\ZZnonneg^{n}$ and $1\le
i\le s$. Note that $\pringshort$ is naturally graded by degree and thus so is
$\pringshort^{s}$. We denote by $\Mon[d][\pringshort^s]$ the set of all
monomials of $\pringshort^s$ of degree $d$.

The \emph{tensor algebra} of a module $\module$ over a ring $\ring$ is denoted
$\tensoralgebra$ and is defined by $\tensoralgebra=\bigoplus_{d=0}^{\infty}
\module^{\otimes d}$. Explicitly, for pure tensors $f_1\otimes\cdots\otimes
f_d\in\module^{\otimes d}$ and $g_1\otimes\cdots\otimes g_e\in\module^{\otimes
e}$ of ranks $d$ and $e$ respectively, we have 
\[
	(f_1\otimes \cdots \otimes f_d)\cdot(g_1\otimes\cdots \otimes g_e)=f_1\otimes\cdots\otimes f_d\otimes g_1\otimes\cdots\otimes g_e\in\module^{\otimes (d+e)}
.\] 
The algebra $\tensoralgebra$ carries a natural grading as a ring, wherein its
homogeneous part of degree $d$ is precisely the $\ring$-module
$\module^{\otimes d}$. 
\begin{proposition}{\cite[Cor.\,A2.3]{Eisenbud1995}}\label{prop:prelim:tensoralg-basis}
	Let $\ring$ be a ring and let $\module$ be a finite free $\ring$-module
	with basis $e_1,\dots, e_s$. Then for any integer $d\ge 1$,
	$\module^{\otimes d}$ is a free module of rank $s^{d}$ and the set
	$\{e_{i_1}\otimes\cdots\otimes e_{i_d}:1\le i_1,\dots,i_d\le s\}$ is an
	$\ring$-basis for $\module^{\otimes d}$.
\end{proposition}

The \emph{symmetric algebra} $\Sym(\module)$ of a module $\module$ over a ring
$\ring$ is simply the quotient $\tensoralgebra/\genby{u\otimes v-v\otimes
u:u,v\in\module}$. The grading on $\tensoralgebra$ naturally induces a grading
on  $\Sym(\module)$, wherein the homogeneous part of degree $d$ of
$\Sym(\module)$ is called the \emph{$d$-th symmetric power of $\module$} and is
denoted $\Sym_d(\module)$.

The \emph{exterior algebra} of a module $\module$ over a ring $\ring$ is
denoted $\extalg$ and is defined by $\extalg=\tensoralgebra/\langle x\otimes
x:x\in\module\rangle$. We denote by $f_1\wedge\cdots\wedge f_d$ the image of
the pure tensor $f_1\otimes\cdots\otimes f_d$ in $\extalg$. The grading on
$\tensoralgebra$ described above naturally induces a grading on $\extalg$. In
this case, the homogeneous part of degree $d$ of $\extalg$ is called the
\emph{d-th exterior power of $\module$} and is denoted $\extalg[d]$. As in the
case of $\module^{\otimes d}$, the abelian group $\extalg[d]$ naturally carries
the structure of an $\ring$-module. 
\begin{proposition}{\cite[Cor.\,A2.3]{Eisenbud1995}}\label{prop:prelim:extalg-basis}
	Let $\ring$ be a ring and let $\module$ be a finite free $\ring$-module
	with basis $e_1,\dots, e_s$. Then for any integer $d\ge 1$, $\extalg[d]$ is a free
	module of rank $\binom{s}{d}$ and the set $\{e_{i_1}\otimes\cdots\otimes
	e_{i_d}:1\le i_1<\cdots<i_d\le s\}$ is an $\ring$-basis for
	$\extalg[d]$.
\end{proposition}

For a module $\module$ over a ring $\ring$, we denote by
$\dual{\module}=\Hom(\module,\ring)$ the \emph{dual module} of $\module$.
A sequence $(f_1,\dots, f_s)\subseteq\ring$ is said to be \emph{$\module$-regular} if
$f_1$ is not a zero-divisor in $\module$ and,
for all $2\le i\le s$, $f_i$ is not a zero-divisor in $\module/\langle f_1,\dots,f_{i-1}\rangle$.
If $\ideal$ is an ideal of $\ring$, the \emph{grade of $\ideal$ with respect to
$\module$}, denoted $\grade(\ideal,\module)$ is the length of a maximal
$\module$-regular sequence of elements of $\ideal$. We take
$\grade(\ideal)=\grade(\ideal,\ring)$.

\myparagraph{Hilbert functions}
For a graded module $\module$ over $\pringshort$ equipped with its natural
grading by degree, the \emph{Hilbert function} of $\module$ is defined by
$\HF_\module(d)=\dim_\field(\module_d)$. The \emph{Hilbert series}
$H_\module(t)=\sum_{d\ge 0}\HF_\module(d)t^d\in\ZZ\llbracket t\rrbracket$ of
$\module$ is the generating function of $\HF_\module(d)$.
\begin{theorem}{\cite[Thm.\,1.1]{Eisenbud1995}}\label{thm:prelim:hilbert-polynomial}
	If $\module$ is a finitely generated graded module over $\pringshort$,
	then $\HF_\module(d)$ is, for sufficiently large $d$, a polynomial
	$P_\module(d)$ of degree at most $\nvars-1$.
\end{theorem}
Pursuant to \cref{thm:prelim:hilbert-polynomial}, the polynomial $P_\module(d)$
is called the \emph{Hilbert polynomial} of  $d$. The \emph{Hilbert regularity}
of $\module$, is the smallest integer $d$ such that for all $d'\ge d$,
$\HF_\module(d')=P_\module(d')$.

\myparagraph{Syzygies and free resolutions}
Free resolutions are a fundamental construction, with many general properties
\cite[Part\,III]{Eisenbud1995} \cite[Ch.\,6]{CoxLittleOShea2007}.
Again, we recall below only what we need for our purposes.

Let $\ring$ be a ring and $\module$ a finite $\ring$-module. An exact sequence
\[
	\cdots\xrightarrow{\partial_{j+1}}\freeresmod_j\xrightarrow{\partial_j}\cdots\xrightarrow{\partial_2}\freeresmod_1\xrightarrow{\partial_1}\freeresmod_0\xrightarrow{\epsilon}\module\to 0
\] 
is a \emph{left resolution} of $\module$. The maps $\partial_i$ are
\emph{boundary homomorphisms}, and the map $\epsilon$ is an \emph{augmentation
homomorphism}. If for each $i$, the module $\freeresmod_i$ is free, then the
resolution is a \emph{free resolution}. For the sake of brevity, we will often
refer to a resolution as above simply by
$(\freeresmod_\bullet\xrightarrow{\epsilon}\module, \partial_\bullet)$. We call
$\sup\{i\in\ZZ:\freeresmod_i\ne 0\}$ the \emph{length} of the resolution
$(\freeresmod_\bullet\xrightarrow{\epsilon}\module, \partial_\bullet)$. The
length of $(\freeresmod_\bullet\xrightarrow{\epsilon}\module,
\partial_\bullet)$ could be infinite and free resolutions of finite length are
\emph{finite free resolutions}.

\begin{theorem}[Hilbert's syzygy theorem, {\cite[Cor.\,19.7]{Eisenbud1995}}]
	Let $\module$ be a finitely generated $\pring$-module. There exists a
	free resolution $(\freeresmod_\bullet\xrightarrow{\epsilon}\module,
	\partial_\bullet)$ of length at most $\nvars$.
\end{theorem}

When $\ring$ is graded and $\module$ is a graded $\ring$-module, $\module$
possesses a free resolution $(\freeresmod_\bullet\xrightarrow{\epsilon}\module,
\partial_\bullet)$ where each $\freeresmod_i$ is graded so that the boundary
maps $\partial_i$ and the augmentation map $\epsilon$ are graded $\ring$-module
homomorphisms. Such free resolutions are called \emph{graded free resolutions}.
Graded free resolutions $(\freeresmod_\bullet\xrightarrow{\epsilon}\module,
\partial_\bullet)$ such that the ranks of each of the $\freeresmod_i$ are
minimal are \emph{minimal free resolutions}.

Let $\ring$ be a ring and $F=(f_1,\dots,f_s)\subseteq\ring$ a sequence of
elements of $\ring$. We define the \emph{syzygy module} of $F$ to be the
$\ring$-module
\[
	\Syz(F)=\{(g_1,\dots,g_s)\in\ring^{s}:g_1f_1+\dots+g_sf_s=0\}
.\] 
If $(\freeresmod_\bullet\xrightarrow{\epsilon}\module,\partial_\bullet)$ is a
free resolution of length $\ell$, with $\rank(\freeresmod_i)=r_i$ then
$\Syz(\epsilon(e_1),\dots,\epsilon(e_{r_0}))=\ker(\epsilon)=\im(\partial_1)$
and for each $1\le i\le\ell$,
$\Syz(\partial_i(e_1),\dots,\partial_i(e_{r_i}))=\ker(\partial_i)=\im(\partial_{i+1})$.

The following consequence of Hilbert's syzygy theorem elucidates the connection
between free resolutions and Hilbert series.

\begin{corollary}{{\cite[Thm.\,4.4]{CoxLittleOShea2005}}}\label{cor:prelim:hilb-free-res}
	Let $\ring$ be a graded ring, let $\module$ be a finitely generated
	graded $\ring$-module, and let
	$(\freeresmod_\bullet\xrightarrow{\epsilon}\module, \partial_\bullet)$
	be a finite graded free resolution of $\module$ of length $\ell$. For
	any $1\le i\le\ell$, let $s_i=\rank(\freeresmod_i)$ and write
	$\freeresmod_i=\bigoplus_{j=1}^{s_i}\ring(-d_i^{(j)})$. Then
	\[
		\HF_\module(d)=\textstyle\sum_{i=0}^{\ell}(-1)^{i}\sum_{j=1}^{s_i}\binom{k+d-d_i^{(j)}-1}{k-1}
	.\] 
\end{corollary}

\myparagraph{Genericity}
\label{subsec:prelim:genericity}
Several of our results rely on genericity assumptions. Let
$\Mon[d][\pringshort]$ be the set of monomials in $\pringshort$ of degree $d$.
For $\nvars,d\in\ZZpos$, and a set
$\frakc=\{\frakc_\tau:\tau\in\Mon[d][\pringshort]\}$ of indeterminates, we call
the polynomial
\[
	\genericf{\frakc}{\nvars}{d}
  =\textstyle\sum_{\tau\in\Mon[d][\pringshort]}\frakc_{\tau}\tau
  \quad \in\pringshort{}[\frakc]
\] 
the \emph{generic homogeneous polynomial in $\nvars$ variables of degree $d$}. A point
$c=(c_\tau:\tau\in\Mon[d][\pringshort])\in\affspace{\binom{\nvars+d-1}{\nvars-1}}$
defines a map
\[
	\phi_\frakc:\pringshort{}[\frakc] \to \pringshort ;
	\quad \frakc_\tau \mapsto c_\tau
\] 
which maps $\genericf{\frakc}{\nvars}{d}$ to a homogeneous polynomial of degree
$d$.

Let $(d_1,\dots,d_s)\in\ZZpos^s$ and let $\frakc^{(1)},\dots,\frakc^{(s)}$ be
sets of indeterminates, with
$\frakc^{(i)}=\{\frakc^{(i)}_\tau:\tau\in\Mon[d_i][\pringshort]\}$ for each
$1\le i\le s$. For a point
$c=(c^{(1)},\dots,c^{(s)})\in\prod_{i=1}^{s}\affspace{\binom{\nvars+d_i-1}{\nvars-1}}$,
the map
\begin{equation*}
	\phi_{c}: \pringshort{}[\frakc^{(1)},\dots,\frakc^{(s)}] \to \pringshort ;
	\quad \frakc^{(i)}_\tau \mapsto c^{(i)}_\tau
\end{equation*}
defines a sequence of polynomials
$(\phi_c(\genericf{\frakc^{(1)}}{\nvars}{d_1}),\dots,\phi_c(\genericf{\frakc^{(s)}}{\nvars}{d_s}))$,
with $\phi_c(\genericf{\frakc^{(1)}}{\nvars}{d_i})$ homogeneous of
degree $d_i$, for each $1\le i\le s$. Given such a point $c$, we will simply
denote by $\specialseq{c}{\nvars}{d_1,\dots,d_s}$ the sequence of polynomials
defined by $c$ in this way.

Similarly, let $\nrows,\ncols\in\ZZpos$ with $\ncols\ge\nrows$ and for $1\le
i\le\nrows$, $1\le j\le\ncols$, let $d_{i,j}\in\ZZpos$ and let
$\frakc^{(i,j)}=\{\frakc^{(i,j)}_\tau:\tau\in\Mon[d_{(i,j)}][\pringshort]\}$ be
a set of indeterminates. For a sequence of points
$c=(c^{(1,1)},\dots,c^{(\nrows,\ncols)})$ with
$c^{(i,j)}\in\affspace{\binom{\nvars+d_{i,j}-1}{\nvars-1}}$, the map
\begin{equation*}
	\phi_c:\pringshort{}[\frakc^{(1,1)},\dots,\frakc^{(\nrows,\ncols)}] \to \pringshort ;
	\quad \frakc^{(i,j)}_\tau \mapsto c^{(i,j)}_\tau
\end{equation*}
defines a matrix
$(\phi_c(\genericf{\frakc^{(i,j)}}{\nvars}{d_{i,j}}))_{i,j}\in\pringshort^{\nrows\times\ncols}$.
Again, given such a sequence of points
$c=(c^{(1,1)},\dots,c^{(\nrows,\ncols)})$, we will simply denote by
$\specialmat{c}{\nvars}{d_{i,j}}\in\pringshort^{\nrows\times\ncols}$ the
$\nrows\times\ncols$ matrix defined by  $c$ in this way. 

The following important fact is what will allow us to use the Eagon-Northcott
complex to compute syzygies amongst maximal minors of polynomial matrices.
\begin{proposition}[{\cite[Thm.\,2.5]{BrunsVetter1988}}]
	Let $\nvars,\nrows,\ncols,d_0\in\ZZpos$ with $\ncols\ge\nrows$. Then
	there exists a Zariski open subset
	$U\subseteq\affspace{\nrows\ncols\binom{\nvars+d_0-1}{\nvars-1}}$ such
	that for all $c\in U$,
	$\grade(\maxminorsidealSmallPar{\specialmat{c}{\nvars}{d_0}})=\ncols-\nrows+1$. 
\end{proposition}

\section{Signature Gr\"obner bases}\label{sec:grobner}

From here on, we take $\ord$ to be the \emph{graded reverse lexicographic} (or
grevlex) order on $\pringshort$, and $\ordpot$ to be the corresponding
\emph{position over term} (or POT) order on $\pringshort^s$ (see e.g.\
\cite[Def.\,2.4, p211]{CoxLittleOShea2005}).

\myparagraph{Gr\"obner bases and modules}
By \cref{prop:prelim:extalg-basis}, if $\module$ is a free
$\pringshort$-module of rank $s$, then $\extalg[d]$ is also a free
$\pringshort$-module of rank $\binom{s}{d}$. Since the basis we fix on
$\extalg[d]$ is not indexed by the integers $1,\dots,\binom{s}{d}$ we slightly
generalize the definition of the POT order: for
$\alpha,\beta\in\ZZnonneg^{\nvars}$ and two strictly increasing sequences $1\le
i_1<\cdots<i_d\le s$, $1\le i_1'<\cdots<i_d'\le s$, we take
$x^{\alpha}(e_{i_1}\otimes\cdots\otimes e_{i_d})\ordpot
x^{\beta}(e_{i_1'}\otimes\cdots\otimes e_{i_d'})$ if and only if
$(i_1,\dots,i_d)\lexgreater(i_1',\dots,i_d')$ or
$(i_1,\dots,i_d)=(i_1',\dots,i_d')$ and $x^{\alpha}\ord x^{\beta}$.

The set of all monomials of $\pringshort$ (resp. $\pringshort^s$) forms a basis
for $\pringshort$ (resp. $\pringshort^s$) as an infinite-dimensional
$\field$-vector space. The \emph{leading monomial} of an element
$f\in\pringshort$ (resp. $f\in\pringshort^s$), denoted $\LM(f)$ (resp.
$\LMmod(f)$) is the largest monomial, with respect to $\ord$ (resp. $\ordpot$),
which appears in the unique representation of $f$ in this $\field$-basis. We
naturally extend the leading monomial notation to sets: for a set
$F\subseteq\pringshort^s$, $\LMmod(F)=\{\LMmod(f):f\in F\}$.

For some $s\in\ZZpos$, a \emph{$\ordpot$-Gr\"obner basis} of a submodule
$\module\subseteq\pringshort^s$ is a set $G\subseteq\module$ such that
$\modgenby{\LMmod(G)}=\modgenby{\LMmod(\module)}$. When $s=1$ so that $\ordpot$ coincides
with $\ord$ and $\module$ is an ideal of $\pringshort$, we call a
$\ordpot$-Gr\"obner basis of $\module$ a $\ord$-Gr\"obner basis.

\myparagraph{Macaulay matrices}
For integers $s,\nvars\in\ZZpos$ and a set
$\modF=\{\modf_1,\dots,\modf_t\}\subseteq\pringshort^s$ of homogeneous
elements, the \emph{Macaulay matrix of $\modF$ in degree $d$ with respect to
$\ordpot$}, denoted $\macmat_d(\modF)$, is constructed as follows: its rows are
indexed by the set $\bigcup_{i=1}^{s}\{\tau
e_i:\tau\in\Mon[d-\deg\modf_i][\pringshort]\}$, its columns are indexed by
$\Mon[d][\pringshort^s]$, ordered decreasingly by $\ordpot$, and for some $1\le
i,j\le s$ and $\tau\in\Mon[d-\deg\modf_i][\pringshort]$,
$\sigma\in\Mon[d][\pringshort^s]$, the entry of the row indexed by $\tau e_i$
in the column indexed by $\sigma e_j$ is the coefficient of $\sigma e_j$ in
$\tau\modf_i$. The monomial $\tau e_i$ is the \emph{signature} of the row of
$\macmat_d(\modF)$ which it indexes.

For $1\le i\le t$, we abbreviate
$\macmat_{d,i}(\modF)=\macmat_d(\{\modf_1,\dots,\modf_i\})$.

A \emph{valid elementary row operation} on a Macaulay matrix $\macmat_d(\modF)$
consists in adding to a row of $\macmat_d(\modF)$ with signature $\tau e_i$ a
$\field$-multiple of a row with some signature $\sigma e_j$, where $\tau
e_i\ordpot\sigma e_j$. Finally, we denote by $\macmatred_d(\modF)$ a
row-echelon form of $\macmat_d$ computed via a sequence of valid elementary row
operations.

Each row of $\macmat_d(\modF)$ can be interpreted as an element of
$\pringshort^s$ by multiplying the entry in a given column by the monomial
which indexes that column and taking the sum over all columns. We refer to rows
of $\macmat_d(\modF)$ (resp. $\macmatred_d(\modF)$) as elements of
$\pringshort^s$, denoting them by $\rows(\macmat_d(\modF))$ (resp.
$\rows(\macmatred_d(\modF))$).

For some $D\in\ZZpos$, we call
\emph{$(D,\ordpot)$-Gr\"obner basis} of $\genby{\modF}$ the union
of the sets \(\rows(\macmatred_d(\modF))\) for
\(d=\{\min_{1\le i\le t}\{\deg\modf_i\},\ldots,D\}\).
This is justified by
the following.
\begin{proposition}{\cite[Sec.\,3]{Lazard1983}}\label{prop:prelim:lazard}
	Let $s,\nvars\in\ZZpos$ and let
	$\modF=\{\modf_1,\dots,\modf_t\}\subseteq\pringshort^s$ be homogeneous
	elements with respect to the standard grading on $\pringshort^s$. Then
	there exists $D\in\ZZpos$ such that a $(D,\ordpot)$-Gr\"obner basis of
	$\genby{\modF}$ is a $\ordpot$-Gr\"obner basis of $\genby{\modF}$. 
\end{proposition}

Moreover, it is shown in \cite[Sec.\,3]{Lazard1983} that generically (in the
sense of \cref{subsec:prelim:genericity}), the integer $D$ in
\cref{prop:prelim:lazard} satisfies the bound $D\le
1+\sum_{i=1}^{t}(\deg(\modf_i)-1)$.

\myparagraph{The matrix-\texorpdfstring{$F_5$}{F5} algorithm}
\cref{prop:prelim:lazard} leads to an algorithm to compute Gr\"obner bases
using linear algebra. This algorithm, known as Lazard's algorithm, is described
in \cite{Lazard1983}. Informally, given a polynomial system
$F\subseteq\pringshort$ and a degree bound $D\in\ZZpos$, it works by building
the matrices $\macmat_d(F)$ and computing from them $\macmatred_d(F)$, for each
degree $\min_{1\le i\le t}\{\deg(f_i)\}\le d\le D$. 

The following proposition, known as the syzygy criterion, lies at the core of
the $\SigGB$ algorithm, which improves upon Lazard's algorithm by building
Macaulay matrices with fewer rows.

\begin{proposition}[Syzygy Criterion, {\cite[Lem.\,6.4]{EderFaugere2016}}]\label{prop:syz-crit}
	Let $s\in\ZZpos$, $F=(\modf_1,\dots,\modf_t)\subseteq\pringshort^s$ be
	homogeneous elements and let $h=(h_1,\dots, h_\ell)$ be a homogeneous
	syzygy of $F$ with $\LMmod(h)=\tau e_i$. 
	\begin{enumerate}[leftmargin=0.5cm]
		\item \label{it:syz-crit-1} The row of $\macmat_{\deg\tau+d_i}$
			with signature $\tau e_i$ is a linear combination of
			rows of $\macmat_{\deg\tau+\deg\modf_i}$ of smaller
			signature.
		\item \label{it:syz-crit-2} For any monomial
			$\sigma\in\pringshort$, the row of
			$\macmat_{\deg\tau+\deg\sigma+\deg\modf_i}$ with
			signature $\sigma\tau e_i$ is a linear combination of
			rows of $\macmat_{\deg\tau+\deg\sigma+\deg\modf_i}$ of
			smaller signature.
	\end{enumerate}
\end{proposition}

Suppose now that $F=(f_1,\dots,f_t)\subseteq\pringshort$ is a polynomial
system. Then for each $1\le i,j\le t$, $f_ie_j-f_je_i\in\Syz(F)$.
Syzygies of this form are called \emph{Koszul syzygies} and the $\SigGB$
algorithm exploits precisely these syzygies to improve upon Lazard's algorithm.

\begin{theorem}[$F_5$ Criterion, {\cite[Thm.\,1]{Faugere2002}}]\label{prop:f5-crit}
	Let $F=(f_1,\dots,f_\ell)$ be a polynomial system in $\pringshort$.
	Then for any $d\in\ZZpos$, any $1\le i\le \ell$, any
	$\tau\in\LM(\rows(\macmatred_{d,i}(F)))$, and any $i<j\le \ell$, the
	row of $\macmat_{d+\deg(f_j)}(F)$ with signature $\tau e_j$ is a linear
	combination of rows of $\macmat_{d+\deg(f_j)}(F)$ with smaller
	signature.
\end{theorem}

We recall here \cite[Algorithm\,1]{GoNeSa23}, which is a slightly modified
version of the standard $\SigGB$ algorithm \cite{BardetFaugereSalvy2015} (see
also \cite[Sec.\,3]{EderFaugere2016}) permitting the input of precomputed
syzygies.

\begin{algorithm}[ht]
	\caption{$\SigGB$\((F,D,S)\)}
	\label{alg:mF5-syz}
  \begin{algorithmic}[1]
	  \Require{A sequence $F=(f_1,\dots, f_t)$ of homogeneous elements of
	  degrees $d_1\le\cdots\le d_t$ in \(\pring^{s}\); a degree bound $D$;
          a set $S$ of syzygies of $F$.}
	  \Ensure{A $(D,\ordpot)$-Gr\"obner basis for $\genby{F}$.}

	  \State \InlineFor{$i$ from $1$ to $t$}{$G_i\gets\emptyset$}
	  \For{$d$ from $d_1$ to $D$}
	  \State $\macmat_{d,0}\gets\emptyset$; $\mathrm{Crit}\gets\LMmod(S)$
	  \For{$i$ from $1$ to $t$}
	  \If{$d<d_i$}
	  $\macmat_{d,i}\gets \macmat_{d,i-1}$
	  \ElsIf{$d=d_i$}
          $\macmat_{d,i}\gets$ concatenate the row $f_i$ to $\macmatred_{d,i-1}$ with signature $e_i$
	  \Else
          \State $\macmat_{d,i}\gets\macmatred_{d,i-1}$
	  \If{$s=1$}
	  \For{$\tau\in\LM(\rows(\macmat_{d-d_i,i-1}))$}
		  \State $\mathrm{Crit}\gets\mathrm{Crit}\cup\{\tau e_i\}$
          \EndFor
	  \EndIf
	  \For{$f\in\rows(\macmatred_{d-1,i})\smallsetminus\rows(\macmatred_{d-1,i-1})$}
	  	\State $\tau e_i\gets$ signature of $f$
		\If{$f=0$}
			\For{$j$ from $1$ to $k$}
				\State $\mathrm{Crit}\gets\mathrm{Crit}\cup\{\tau x_j e_i\}$
			\EndFor
	    	\EndIf
	  \EndFor
          	\For{$f\in\rows(\macmat_{d-1,i})\smallsetminus\rows(\macmat_{d-1,i-1})$}
			\State $\tau e_i\gets$ signature of $f$
			\For{$j\in\{\max\{j':x_{j'}\mid\tau\},\dots,k\}$}\label{line:var-select}
				\If{$\tau x_j e_i\notin\mathrm{Crit}$}
				$\macmat_{d,i}\gets$ concatenate the row $x_jf$ to $\macmat_{d,i}$ with signature $\tau x_j e_i$\label{line:mac-mat-add-row}
				\EndIf
			\EndFor
        	\EndFor
	\EndIf
      \State $\macmatred_{d,i} \gets$ reduced row echelon form of $\macmat_{d,i}$ obtained via a sequence of valid elementary row operations
      \State $G_i\gets G_i\cup\{f\in\rows(\macmatred_{d,i}):f\notin \genby{\LMmod(G_i)}\}$ 
    \EndFor
    \EndFor
    \State \Return $G_1,\dots,G_t$
  \end{algorithmic}
\end{algorithm}

\section{The first syzygies of maximal minors}\label{sec:eagonnorthcott}

First defined in \cite{EagonNorthcott1962}, the Eagon-Northcott complex is a
complex of free modules associated to a matrix with entries in any commutative
ring with unity. We are specifically concerned with the first syzygies of
maximal minors of some polynomial matrix. The Eagon-Northcott complex provides
access to them.

\subsection{The Eagon-Northcott complex}\label{subsec:en:eagon-northcott}

\begin{theorem}[{\cite[Thm.\,1]{EagonNorthcott1962}, \cite[Thm.\,A2.60]{Eisenbud2005}}]\label{thm:en:EN}
	Let $\ring$ be a commutative ring with unity and let $\polymat$ be a
	$\nrows\times \ncols$ matrix with entries in $\ring$, with $
	\nrows\le\ncols$. For each $0\le i\le \ncols-\nrows$, let
	\[
		\freeresmod_i=\dual{\Sym_i\ring^{\nrows}}\otimes\extalg[\nrows+i][\ring^{\ncols}]
	.\] 
	Then there are graded morphisms
	$\partial_i:\freeresmod_i\to\freeresmod_{i-1}$, $1 \le i \le \ncols-\nrows$, such that the complex
	$\EN(\polymat)=(\freeresmod_\bullet\xrightarrow{\epsilon}\ring/\maxminorsideal{\polymat},\partial_\bullet)$
	is a free resolution if and only if $\grade(\maxminorsideal{\polymat})=\ncols-\nrows+1$.
\end{theorem}

We make explicit the first boundary morphism $\partial_1$, whose image is
precisely the first syzygy module of $\maxminorsideal{\polymat}$. First, by the
definition of the free modules $\freeresmod_i$, the map $\partial_1$ is a map
\[
	\partial_1:\dual{\ring^{\nrows}}\otimes\extalg[\nrows+1][\ring^{\ncols}]\to\extalg[\nrows][\ring^{\ncols}]
.\] 
We take as a basis for $\dual{\ring^{\nrows}}$ the standard basis functionals
$e_i$ for $1\le i\le \ncols$. It follows immediately from
\cref{prop:prelim:tensoralg-basis,prop:prelim:extalg-basis} that a basis for
the $\ring$-module
$\dual{\ring^{\nrows}}\otimes\extalg[\nrows+1][\ring^{\ncols}]$ is given by
\[
	\{e_i\otimes(e_{i_1}\wedge\cdots\wedge e_{i_\nrows}):1\le i\le \nrows, 1\le
	i_1<\cdots<i_\nrows\le\ncols\}
.\] 
Subsequently, $\partial_1(e_i\otimes (e_{i_1}\wedge\cdots\wedge
e_{i_{\nrows+1}}))$ equals
\[
	\sum_{t=1}^{\nrows+1}(-1)^{t-1}\left(e_{i_t}\polymat^{T}e_i\right)(e_{i_1}\wedge\cdots\wedge
	\widehat{e_{i_t}}\wedge\cdots\wedge e_{i_{\nrows+1}})
.\] 
The map
$\epsilon:\extalg[\nrows][\ring^{\ncols}]\to\ring/\maxminorsideal{\polymat}$ is
given by
\[
	\epsilon(e_{i_1}\wedge\cdots\wedge e_{i_\nrows})=\det([1\cdots p\mid i_1\cdots i_\nrows]_A)
.\] 

The image of $\partial_1$ (and thus $\Syz(\maxminorssystem{\polymat})$ then
admits an explicit description. For each $\nrows\times (\nrows+1)$ submatrix
$\polymat'$ of $\polymat$, the determinant of the square matrix formed by
duplicating any row of $\polymat'$, computed via Laplace expansion around the
duplicated row, is zero. 

\begin{example}\label{ex:en:duplication}
	Let $\nrows=2$, $\ncols=4$, and suppose
	 \[
		 \polymat=
		 \begin{pmatrix} 
			f_{11} & f_{12} & f_{13} & f_{14}\\
			f_{21} & f_{22} & f_{23} & f_{24}
		 \end{pmatrix}\in\ring^{2\times 4}
	.\] 
	We have 
	\[
		\partial_1(e_1\otimes (e_2\wedge e_3\wedge
		e_4))=f_{12}(e_3\wedge e_4)-f_{13}(e_2\wedge
		e_4)+f_{14}(e_2\wedge e_3)
	.\] 
	Note that the fact that $\partial_1(e_1\otimes (e_2\wedge e_3\wedge
	e_4))\in\ker\epsilon$ is precisely the statement that the determinant
	\[
		\det
			\begin{pmatrix} 
				f_{12} & f_{13} & f_{14}\\
				f_{12} & f_{13} & f_{14}\\
				f_{22} & f_{23} & f_{24}
			\end{pmatrix}
	\] 
  is zero.
	More explicitly, $\epsilon(\partial_1(e_1\otimes (e_2\wedge e_3\wedge
	e_4)))$ is simply the determinant of this matrix, computed via
	Laplace expansion along its first row. Analogously,
	$\epsilon(\partial_1(e_2\otimes (e_1\wedge e_3\wedge e_4)))$ is simply
	the determinant of the singular matrix
	\[
		\begin{pmatrix} 
			f_{21} & f_{23} & f_{24}\\
			f_{11} & f_{13} & f_{14}\\
			f_{21} & f_{23} & f_{24}
		\end{pmatrix} 
	\]
	computed via Laplace expansion along its first row.
\end{example}

\subsection{Leading terms of syzygies}

We start with a consequence of the description of the first syzygy module
provided by the Eagon-Northcott complex.

\begin{proposition}\label{prop:en:columns}
	Let $\polymat=(\polymatentry_{i,j})$ be an $\nrows\times\ncols$ matrix with
  entries in $\pring$, with $\nrows\le\ncols$. For each $1\le k\le
  \ncols-\nrows$, let $\idealJ_k(\polymat)$ be the ideal $\langle
  \polymatentry_{i,j}:1\le i\le\nrows, j\le k\rangle$ of $\pring$. Let
  $\mathscr{H}$ be the set
  \[\bigcup_{k=1}^{\ncols-\nrows}\bigcup_{k+1<i_2<\cdots<i_\nrows\le\ncols}\left\lbrace\LM(g)(e_{k+1}\wedge e_{i_2}\wedge\cdots
      \wedge e_{i_\nrows}):g\in \idealJ_k(\polymat)\right\rbrace.\] Then the
  module $\modgenby{\mathscr{H}}$ is a submodule of
  $\LMmod(\Syz(\maxminorssystem{\polymat}))$.
\end{proposition}

\begin{proof} 
	Fix $1\le k\le\ncols-\nrows$. Let $i_2,\dots,i_\nrows\in\ZZpos$ be
	integers such that $k+1<i_2<\cdots<i_\nrows<\ncols$, and let
	$g\in\idealJ_k(\polymat)$. Then there exist polynomials
	$h_{i,j}\in\pringshort$ such that $g=\sum_{i,j}h_{i,j}a_{i,j}$. Let
	\[
		G=\sum_{i,j}h_{i,j}\partial_1(e_i\otimes(e_j\wedge e_{k+1}\wedge e_{i_2}\wedge\cdots\wedge e_{i_\nrows}))
	.\] 
	We claim that
	\[
		\LMmod(G)=\LM(g)(e_{k+1}\wedge e_{i_2}\wedge \cdots\wedge e_{i_\nrows})
	.\] 
	Now taking $\phi=\dual{(e_{k+1}\wedge\cdots\wedge e_{i_\nrows})}$, we have
	\begin{align*}
		\phi(G)&=\sum_{i,j}\phi(h_{i,j}\partial_1(e_i\otimes (e_j\wedge e_{k+1}\wedge e_{i_2}\wedge\cdots\wedge e_{i_\nrows})))\\
		       &=\sum_{i,j}h_{i,j}\phi(\partial_1(e_i\otimes (e_j\wedge e_{k+1}\wedge e_{i_2}\wedge\cdots\wedge e_{i_\nrows})))\\
																					&=\sum_{i,j}h_{i,j}a_{i,j} = g
	\end{align*}
	By the definition of $\partial_1$, only those basis vectors of
	$\extalg[\nrows][\pringshort^\ncols]$ of the form
	$e_{j}\wedge\cdots\wedge\widehat{e_{i_t}}\wedge\cdots\wedge
	e_{i_\nrows}$ appear with nonzero coefficient in $G$. The largest of
	these basis vectors with respect to the lexicographic order is clearly
	$e_{k+1}\wedge e_{i_2}\wedge\cdots\wedge e_{i_\nrows}$, so our claim is
	proven. Having constructed an element of
	$\LMmod(\Syz(\maxminorssystem{\polymat}))$ whose leading term is
	precisely $\LM(g)(e_{k+1}\wedge e_{i_2}\wedge\cdots\wedge
	e_{i_\nrows})$, we are done.
\end{proof}

\begin{remark}
	We have seen, in \cref{subsec:en:eagon-northcott}, that the syzygies
	described by the Eagon-Northcott complex between the maximal minors of
	a $\nrows\times\ncols$ matrix $\polymat$ over $\pringshort$ are given
	by choosing a $\nrows\times(\nrows+1)$ submatrix $\polymat'$ of
	$\polymat$, duplicating any row of $\polymat'$, and computing the
	determinant of this matrix by Laplace expansion over the duplicated row
	(see \cref{ex:en:duplication} for an example). The $\ordpot$-leading
	term of such a syzygy is simply the leading term of the leftmost entry
	of the duplicated row.

	Upon fixing a maximal minor of $\polymat$, the leading monomials of a
	$\ord$-Gr\"obner basis of the polynomial system formed by the set of
	columns to the left of the leftmost column of this minor are leading
	monomials of syzygies amongst the maximal minors of $\polymat$.
\end{remark}

\cref{prop:en:columns} leads directly to the following algorithm.

\begin{algorithm}[ht]
	\caption{$\MaxMinorsSigGB(\polymat, D)$}
	\label{alg:maxminorsgb}
	\begin{algorithmic}[1]
		\Require{A matrix
		$\polymat=(\polymatentry_{i,j})\in\pring^{\nrows\times\ncols}$
		of homogeneous polynomials, with $\nrows\le\ncols$ and an integer $D$.}
		\Ensure{A $(D,\ord)$-Gr\"obner basis of
		$\maxminorsideal{\polymat}$.} 
		\State $C\gets \{a_{1,1},\dots,a_{\nrows,1},\dots,a_{1,\ncols-\nrows},\dots,a_{\nrows,\ncols-\nrows}\}$
		\State $G_1,\dots,G_{\nrows(\ncols-\nrows)}\gets \SigGB(C,\emptyset,D-\min_{f\in \detsystem{\nrows}{\polymat}}\{\deg(f)\})$ 
		\State $H\gets\emptyset$
		\For{$i\in \{1,\dots,\nrows(\ncols-\nrows)\}$}
		\For{$f\in\detsystem{\nrows}{\left[1\cdots\nrows\mid\left(\left\lfloor\frac{i}{\nrows}\right\rfloor+1\right)\cdots \ncols\right]_\polymat}$}
		\State $j\gets$ index of $f$ in $\detsystem{\nrows}{\polymat}$
		\State $H\gets H\cup\{\LM(g)e_j:g\in G_i\}$
		\EndFor
		\EndFor
		\State \Return $\SigGB(\detsystem{\nrows}{\polymat},H,D)$ \label{line:maxminorsgb:finalgb}
	\end{algorithmic}
\end{algorithm}

\begin{theorem}\label{thm:en:alg}
	Algorithm~$\MaxMinorsSigGB$ is correct.
\end{theorem}
\begin{proof}
	This follows from the correctness of $\SigGB$
  \cite[Thm.\,9]{BardetFaugereSalvy2015},
  and from
  \cref{prop:en:columns} which establishes that the set $H$ input to $\SigGB$ on
  \cref{line:maxminorsgb:finalgb} is a subset of
  $\LMmod(\Syz(\maxminorssystem{\polymat}))$.
\end{proof}

\section{Critical points}\label{sec:criticalpoints}

Recall that for a set of homogeneous polynomials
$F=(f_1,\dots,f_\nrows)\subseteq\pringshort$, and a homogeneous polynomial
$g\in\pringshort$, our goal is to compute a Gr\"obner basis for
$\critideal=\maxminorsideal{\jac(g,F)}+\genby{F}$.  Note that if $g$ and the
$f_i$'s are affine, by \cite[Ch.\,8, Sec.\,4, Thm.\,4]{CoxLittleOShea2007}, one
can simply homogenize them with respect to a variable $h$ which is smaller than
all of the $x_i$, apply the algorithms in this paper, then set $h=1$.

Via a minor modification of \cref{alg:maxminorsgb}, we obtain an algorithm
which computes a Gr\"obner basis for the ideal $\critideal$.

\begin{algorithm}[ht]
  \caption{$\CritGB(F,g,D)$}
	\label{alg:criticalpointsgb}
	\begin{algorithmic}[1]
		\Require{A system of homogeneous polynomials
		$F=(f_1,\dots,f_\nrows)\subseteq\pringshort$, a homogeneous polynomial $g\in\pringshort$,
    and an integer \(D\).}
		\Ensure{A $(D,\succ)$-Gr\"obner basis of
		$\critideal$.}
		\State $J\gets\jac(g,F)$
		\State $C\gets \{J_{1,1},\dots,J_{\nrows+1,1},\dots,J_{1,\nvars-\nrows-1},\dots,J_{\nrows,\nvars-\nrows-1}\}$
		\State $G_1,\dots,G_{(\nrows+1)(\nvars-\nrows-1)}\gets \SigGB\left(C,D-\min\left\{\deg\left(\frac{\partial f_i}{\partial x_j}\right) \right\}\right)$ 
		\State $H\gets\emptyset$
		\For{$i\in \{1,\dots,(\nrows+1)(\nvars-\nrows-1)\}$}
		\For{$f\in\detsystem{\nrows+1}{\left[1\cdots \nrows+1\mid \left(\left\lfloor \frac{i}{\nrows+1} \right\rfloor+1\right)\cdots \nvars\right]_J}$}
		\State $j\gets$ index of $f$ in $\detsystem{\nrows+1}{J}$
		\State $H\gets H\cup\{\LM(g)e_j:g\in G_i\}$
		\EndFor
		\EndFor
		\State \Return $\SigGB(F\cup\detsystem{\nrows+1}{J},H,D)$\label{line:critical:finalgb}
	\end{algorithmic}
\end{algorithm}

\begin{proposition}
  Algorithm~$\CritGB$ is correct.
\end{proposition}
\begin{proof}
	The only modification made to \cref{alg:maxminorsgb} to obtain
	\cref{alg:criticalpointsgb} is to add the set $F$ to the polynomial
	system upon which we run $\SigGB$. Thus, the correctness
	follows immediately from that of \cref{alg:maxminorsgb}, proven in
	\cref{thm:en:alg}.
\end{proof}
For a system of homogeneous polynomials
$F=(f_1,\dots,f_\nrows)\subseteq\pringshort$ and a polynomial
$g\in\pringshort$, there may exist, a priori, nontrivial syzygies between the
polynomials in $F$ and the maximal minors of $\jac(g,F)$. Generically, this
does not occur.

\begin{proposition}\label{prop:criticalpoints:syzygies}
	Let $\nvars,\nrows,d_0\in\ZZpos$. There exists a nonempty Zariski open
	subset $U\subseteq\affspace{(\nrows+1)\binom{\nvars+d_0-1}{\nvars-1}}$
	such that for all $c\in U$, taking
	$(g,f_1,\dots,f_\nrows)=\specialseq{c}{\nvars}{(d_0,\dots,d_0)}$,
	\[
		\Syz(F\cup\maxminorssystem[\nrows+1]{\jac(g,F)})=\Syz(F)\oplus\Syz(\maxminorssystem{\jac(g,F)})
	\] 
	where $F=(f_1,\dots,f_\nrows)$.
\end{proposition}
\begin{proof}
	By \cite[Lem.\,2.2]{Spa14}, there exists a nonempty Zariski open subset
	$U\subseteq\affspace{(\nrows+1)\binom{\nvars+d_0-1}{\nvars-1}}$ such
	that for all $c\in U$, taking
  \[
    (g,f_1,\dots,f_\nrows)=\specialseq{c}{\nvars}{(d_0,\dots,d_0)}
  .\]
  and
	$F=(f_1,\dots,f_\nrows)$, the sequence $(f_1,\dots,f_\nrows)$ is a
	$\pringshort/\maxminorsideal{\jac(g,F)}$-regular sequence. Since for
	such $c$, $F$ is also a regular sequence in $\pringshort$, the two
	$\pringshort$-modules $\pringshort/\maxminorsideal{\jac(g,F)}$ and
	$\pringshort/\genby{F}$ are $\Tor$-independent. That is, for all $i\ge
	1$,
	\[
	\Tor_i^{\pringshort}(\pringshort/\maxminorsideal{\jac(F)},\pringshort/\genby{F})=0
	.\] 
	It follows that $\EN(\jac(F))\otimes_{\pringshort}\koszul(F)$ is a free
	resolution of the tensor product
	$\pringshort/\genby{F}\otimes_{\pringshort}\pringshort/\maxminorsideal{\jac(F)}\cong\pringshort/\left(\genby{F}+\maxminorsideal{\jac(F)}\right)$.
\end{proof}

\section{Complexity analysis}\label{sec:complexity}

The complexity of linear-algebra based Gr\"obner basis algorithms is governed
by the cost of echelonizing Macaulay matrices. The work that we have done thus
far allows us to estimate these costs, since the sizes and ranks of the
Macaulay matrices computed can be deduced from the Eagon-Northcott complex.

\subsection{New complexity bound}

Recall that given a polynomial system $F\subseteq\pringshort$, the columns of
$\macmat_d(F)$, are indexed by the monomials of degree $d$ in $\pringshort$. We
are left to compute the rank of $\macmat_d(F)$ and the number of rows of
$\macmat_d(F)$ taken into account by our algorithm. We first count the number
of syzygies from \cref{prop:en:columns}.

\begin{proposition}\label{prop:analysis:num-syz}
	Let $\polymat=(\polymatentry_{i,j})$ be a $\nrows\times\ncols$ matrix with
  entries homogeneous polynomials of degree $d_0$ in $\pringshort$, with
  $\ncols\ge\nrows$. For each $1\le k\le\ncols-\nrows$, let
  $\idealJ_k(\polymat):=\genby{\polymatentry_{i,j}:1\le i\le\nrows, 1\le j\le
    k}\subseteq\pringshort$. Then for any $D\in\ZZpos$ and any $d\in
  \{d_0\nrows, \ldots, D\}$, the number of elements of degree $d-\nrows d_0$ of
  the set $H$ computed in \cref{alg:maxminorsgb} is
	\[
		\sum_{k=1}^{\ncols-\nrows}\HF_{\idealJ_k(\polymat)}(d-\nrows d_0)\binom{\ncols-k-1}{\nrows-1}
	\] 
\end{proposition}
\begin{proof}
	Let $H_{d-\nrows d_0}\subseteq H$ be the subset of the set $H$ in
  \cref{line:maxminorsgb:finalgb} of \cref{alg:maxminorsgb} consisting of
  elements of degree $d-\nrows d_0$. This set $H$
	is precisely the set $\mathscr{H}$ defined in the statement of
  \cref{prop:en:columns}. We can therefore write  
	\begin{align*}
		\#H_{d-\nrows d_0} &=\sum_{k=1}^{\ncols-\nrows}
      \sum_{k+1<i_2<\dots<i_\nrows\le\ncols}\left\lbrace\LM(g)(e_{k+1}\wedge e_{i_2}\wedge\cdots\wedge e_{i_\nrows}):\right .\\
    &\qquad\qquad\qquad\qquad\qquad \left . g\in\idealJ_k(\polymat),\deg(g)=d-\nrows d_0\right\rbrace\\
		&=\sum_{k=1}^{\ncols-\nrows}
    \sum_{k+1<i_2<\cdots<i_\nrows\le\ncols}\HF_{\idealJ_k(\polymat)}(d-\nrows d_0)
	.\end{align*} 
	The last equality follows from the fact that the number of monomials of
	of $\LM(\idealJ_k(\polymat))$ of degree $d-\nrows d_0$ is
	$\HF_{\idealJ_k(\polymat)}(d-\nrows d_0)$. The result follows from the
	fact that for $1\le k\le\ncols-\nrows$, there are
	$\binom{\ncols-k-1}{\nrows-1}$  sequences of the form
	$k+1<i_2<\cdots<i_\nrows\le\ncols$. 
\end{proof}

Using \cref{prop:analysis:num-syz}, we are left to compute the Hilbert functions
of the ideals $\idealJ_k(\polymat)$ of flattened columns, of course under
certain genericity assumptions. To do this, we rely on Fr\"oberg's conjecture
\cite[Sec.\,1]{Froberg85}, which we reformulate below. 

In what follows, for polynomials $P(t),Q(t)\in\ZZ[t]$, we denote by
$\left[\frac{P(t)}{Q(t)}\right]_+$ the power series expansion of
$\frac{P(t)}{Q(t)}$, truncated at its first non-positive coefficient.

\begin{conjecture}[{\cite[Sec.\,1]{Froberg85},\cite[Conj.~1]{Nicklasson17}}]\label{conj:analysis:froberg}
	Consider $(f_1,\dots,f_m)$ be a sequence of homogeneous polynomials in
	$\pringshort$, whose coefficients are algebraically independent. For each
	$1\le i\le m$, let $d_i=\deg(f_i)$. Then
	\[
		H_{\pringshort/\genby{f_1,\dots,f_m}}(t)=\left[\frac{\prod_{i=1}^{m}(1-t^{d_i})}{(1-t)^{n}}\right]_+
	.\] 
\end{conjecture}

\begin{proposition}\label{prop:analysis:semi-regular-hf}
	Let $m,\nvars\in\ZZpos$ and let $F = (f_1,\dots,f_m)\subseteq\pring$ be a
	sequence of homogeneous polynomials, all of degree $d_0$. Let $D$ be
	the Hilbert regularity of $\genby{F}$. If $F$
	is a semi-regular sequence, and \cref{conj:analysis:froberg} is true,
	then for any $d\ge 0$,
	\[
		\HF_{\pringshort/\genby{f_1,\dots,f_m}}(d)=
		\begin{cases}
			\sum_{j=0}^{\left\lfloor\frac{n+d-1}{d_0}\right\rfloor}(-1)^{j}\binom{\nvars+d-d_0j-1}{\nvars-1}\binom{m}{j} & \text{ if } d<D\\
			0 & \text{ if } d\ge D
		\end{cases}
	.\] 
\end{proposition}
\begin{proof}
	By \cref{conj:analysis:froberg}, the Hilbert series of
	$\pringshort/\genby{F}$ is
	\[
		H_{\pringshort/\genby{f_1,\dots,f_m}}(t)=\left[\frac{(1-t^{d_0})^m}{(1-t)^n}\right]_{+}
	.\] 
The numerator $(1-t^{d_0})^{m}$ can be expanded as $\sum_{j=0}^{m}(-1)^{j}\binom{m}{j}t^{jd_0}$
	while the reciprocal of the denominator has the classical expansion
	\[
		\frac{1}{(1-t)^{n}}=\sum_{j\ge 0}\binom{\nvars+j-1}{\nvars-1}t^{j}
	.\] 
	The result follows by taking the product of these expansions.
\end{proof}

\begin{proposition}\label{prop:analysis:jac-columns-semi-regular}
	Let $\nvars\in\ZZpos$. For any $\nrows\le\nvars$ and any
	$d_0\in\ZZpos$, there exists a Zariski open set
	$U\subseteq\affspace{(\nrows+1)\binom{\nvars+d_0-1}{\nvars-1}}$ such
	that for all $c\in U$, taking
	$(g,f_1,\dots,f_\nrows)=\specialseq{c}{\nvars}{d_0,\dots,d_0}$, the
	sequence
	\[
	\left(\frac{\partial g}{\partial x_1},\frac{\partial f_1}{\partial
	x_1},\dots,\frac{\partial f_\nrows}{\partial x_1},\dots,\frac{\partial
	g}{\partial x_{\nvars-\nrows-1}},\frac{\partial f_1}{\partial
	x_{\nvars-\nrows-1}},\dots,\frac{\partial f_\nrows}{\partial
	x_{\nvars-\nrows-1}}\right)
	\] 
	formed by the leftmost $\nvars-\nrows-1$ columns of
	$\jac(g,f_1,\dots,f_\nrows)$ is semi-regular.
\end{proposition}
\begin{proof}
	Let $\frakc^{(1)},\dots,\frakc^{(\nrows+1)}$ be sets of indeterminates,
	with $\frakc^{(i)}=\{\frakc_\tau^{(i)}:\tau\in\Mon[d_0]$. For any $1\le
	i\le \nrows+1$ and for $1\le j\le\nvars-\nrows-1$, the coefficients of the
	partial derivative
	$\frac{\partial\genericf{\frakc^{(i)}}{\nvars}{d_0}}{\partial x_j}$ are
	polynomials in the indeterminate coefficients $\frakc^{(i)}$.

	For any $d\ge d_0$, (upon fixing bases for the domain and codomain),
	the multiplication map by
	$\frac{\partial\genericf{\frakc^{(i)}}{\nvars}{d_0}}{\partial x_j}$
  \[
  \left(\begin{array}{c}
    \pringshort \\
    \hline 
    \genby{\frac{\partial\genericf{\frakc^{(1)}}{\nvars}{d_0}}{\partial
      x_1},\dots,\frac{\partial\genericf{\frakc^{(i-1)}}{\nvars}{d_0}}{\partial
  x_{j-1}}}
    \end{array}
    \right)_{d-d_0}
    \to
    \left(\begin{array}{c}
        \pringshort \\
        \hline
        \genby{\frac{\partial\genericf{\frakc^{(1)}}{\nvars}{d_0}}{\partial
          x_1},\dots,\frac{\partial\genericf{\frakc^{(i-1)}}{\nvars}{d_0}}{\partial
      x_{j-1}}}
      \end{array}\right)_d
  \]
  	is represented by a matrix whose entries are rational functions in the
  	indeterminate coefficients $\frakc^{(i)}$.

	The points $c\in\affspace{(\nrows+1)\binom{\nvars+d_0-1}{\nvars-1}}$
	such that this map is full-rank form a Zariski open subset. By
	intersecting all such subsets for all $1\le i\le\nrows+1$ and $1\le
	j\le\nvars-\nrows-1$, we obtain the set $U$ we seek.
\end{proof}

\begin{corollary}\label{cor:analysis:columns-hf}
	Let $\nvars\in\ZZpos$ and assume \cref{conj:analysis:froberg} is true.
	For any $\nrows\le\nvars$ and any $d_0\in\ZZpos$, there exists a
	Zariski open set
	$U\subseteq\affspace{(\nrows+1)\binom{\nvars+d_0-1}{\nvars-1}}$ such
	that for all $c\in U$, taking
	$(g,f_1,\dots,f_\nrows)=\specialseq{c}{\nvars}{d_0,\dots,d_0}$, for any
	$1\le k\le\nvars-\nrows-1$, $\HF_{\idealJ_k(\jac(g,F))}(d)$ is 
	\[
		\begin{cases}
			\binom{\nvars+d-1}{\nvars-1}-\sum_{j=0}^{\left\lfloor\frac{n+d-1}{d_0}\right\rfloor}(-1)^{j}\binom{\nvars+d-d_0j-1}{\nvars-1}\binom{(\nrows+1)k}{j} & \text{ if } d<D\\
			\binom{\nvars+d-1}{\nvars-1} & \text{ if } d\ge D
		\end{cases}
	\] 
	where $\idealJ_k(\jac(g,F))
      $ is the ideal generated by the first $k$ columns of $\jac(g,F)$, and $D$ 
      is its Hilbert regularity.
\end{corollary}
\begin{proof}
	Let $U$ be the set defined in
	\cref{prop:analysis:jac-columns-semi-regular} and take any $c\in U$.
	Then for any $1\le k\le\nvars-\nrows-1$, the set of generators for
	$\idealJ_k(\jac(g,F))$ given by the first $k$ columns of $\jac(g,F)$
	forms a semi-regular sequence. We can therefore apply
	\cref{prop:analysis:semi-regular-hf} to obtain the Hilbert function of
	$\pringshort/\idealJ_k(\jac(g,F))$. Since the Hilbert function of
	$\pringshort$ itself is
	$\HF_{\pringshort}(d)=\binom{\nvars+d-1}{\nvars-1}$, the result
	follows.
\end{proof}

\begin{conjecture}\label{conj:analysis:U-nonempty}
	The set $U$ defined in \cref{prop:analysis:semi-regular-hf} is
	nonempty.
\end{conjecture}
\begin{remark}
	Such a conjecture is a variation of Fr\"oberg's.
\end{remark}

Finally, we compute the ranks of the Macaulay matrices associated to the
maximal minors of a polynomial matrix from the Eagon-Northcott complex.

\begin{proposition}\label{prop:analysis:en-hf}
	Let $\polymat$ be a $\nrows\times\ncols$ matrix with entries
	homogeneous polynomials of degree $d_0$ in $\pring$, with
	$\ncols\ge\nrows$. If
	$\grade(\maxminorsideal{\polymat})=\ncols-\nrows+1$, then
	\[
		\HF_{\maxminorsideal{\polymat}}(d)=\sum_{j=0}^{\ncols-\nrows}(-1)^{j}\binom{\nvars+d-(\nrows+j)d_0-1}{\nvars-1}\binom{\nrows+j-1}{\nrows-1}\binom{\ncols}{\nrows+j}
	.\] 
\end{proposition}
\begin{proof}
	Since $\grade(\maxminorsideal{\polymat})=\ncols-\nrows+1$, by
	\cref{thm:en:EN}, the Eagon-Northcott complex is a free resolution of
	$\maxminorsideal{\polymat}$. In order to turn the Eagon-Northcott
	complex into a graded resolution, we need to shift the grading on the
	component free modules to ensure that the boundary homomorphisms are
	graded.

	First, the degree of the maximal minors of $\polymat$ is $\nrows d_0$.
	As such, in order to make the augmentation homomorphism $\epsilon$ of
	$\EN(\polymat)$ graded, we need only replace $\freeresmod_0$ by
	$\freeresmod_0(-\nrows d_0)$.

	Now, by the description of the boundary homomorphisms in
	\cite[A2H]{Eisenbud2005} (see also \cite[Exa.~A2.69]{Eisenbud2005}),
	each boundary homomorphism (except for the augmentation homomorphism)
	can be represented by a matrix with entries linear in the entries of
	$\polymat$. Thus, in order to make these boundary homomorphisms graded,
	we need to replace $\freeresmod_j$ by $\freeresmod_j(-\nrows d_0-id_0)$
	for each $1\le j\le \ncols-\nrows$. 

	Finally, we have $\rank(\freeresmod_0(-\nrows d_0))=\binom{\ncols}{\nrows}$ and 
	\begin{align*}
		\rank(\freeresmod_j(-d_0(\nrows+j)))&=\rank\left(\dual{\Sym_j(\ring^\nrows)}\otimes\extalg[\nrows+j][\ring^\ncols]\right)\\
						    &=\binom{\nrows+j-1}{\nrows-1}\binom{\ncols}{\nrows+j}
	.\end{align*} 
	The result then follows from \cref{cor:prelim:hilb-free-res}.
\end{proof}

\begin{proposition}\label{prop:analysis:crit-hf}
	Let $F=(f_1,\dots,f_\nrows)\subseteq\pring$ be a regular sequence of
	homogeneous polynomials, all of degree $d_0$, and let $g\in\pring$ be a
	homogeneous polynomial of degree $d_0$. Then the Hilbert function of
	the ideal $\critideal=\genby{F}+\maxminorsideal[\nrows+1]{\jac(g,F)}$
	is given by $\HF_{\pringshort/\critideal}(d)$ which is
  \begin{align*}
    &\sum_{i=0}^{\nrows}(-1)^{i}\binom{\nrows}{i}\left(\binom{\nvars+d-id_0-1}{\nvars-1}-\right .\\
    &\left .\sum_{j=0}^{\nvars-\nrows-1}(-1)^{j}\binom{\nvars+d-(\nrows+j+i+1)d_0-1}{\nvars-1}\binom{\nrows+j}{\nrows}\binom{\nvars}{\nrows+j+1}\right)
  \end{align*}
\end{proposition}
\begin{proof}
	By \cite[Lem.~2.2]{Spa14}, the sequence $F$ is a
	$\pringshort/\maxminorsideal[\nrows+1]{\jac(g,F)}$-regular sequence. It follows
	(see e.g. \cite[Exe.~10.13(a)]{Eisenbud1995}) that 
	 \begin{equation}
		 H_{\pringshort/\critideal}(t)=H_{\pringshort/\maxminorsideal[\nrows+1]{\jac(g,F)}}(t)(1-t^{d_0})^{\nrows}\label{eq:analysis:crit-system-hs}
	.\end{equation} 
By \cref{prop:analysis:en-hf},
$\HF_{\maxminorsideal[\nrows+1]{\jac(g,F)}}(d)$ equals
	\begin{equation}
		\sum_{j=0}^{\nvars-\nrows-1}(-1)^{j}\binom{\nvars+d-(\nrows+j+1)d_0-1}{\nvars-1}\binom{\nrows+j}{\nrows}\binom{\nvars}{\nrows+j+1}\label{eq:analysis:crit-jacobian-hf}
	.\end{equation}
	Since the Hilbert series
	$H_{\pringshort/\maxminorsideal[\nrows+1]{\jac(g,F)}}(t)$ is the
	generating series of the difference between the Hilbert function of
	$\pringshort$ (which is simply
	$\HF_{\pringshort}(d)=\binom{\nvars+d-1}{\nvars-1}$) and this Hilbert
	function, the result follows by combining
	\cref{eq:analysis:crit-jacobian-hf} with
	\cref{eq:analysis:crit-system-hs} and expanding.
\end{proof}

We use the Hilbert functions to estimate the cost of echelonizing each of the
Macaulay matrices encountered in \cref{alg:criticalpointsgb}.

\begin{theorem}\label{thm:analysis:final-bound}
	Let $F=(f_1,\dots,f_\nrows)\subseteq\pring$ be a regular sequence of
	homogeneous polynomials, all of degree $d_0$, and let $g\in\pring$ be a
	homogeneous polynomial of degree $d_0$. Finally, for each $1\le
	k\le\nvars-\nrows-1$, let $\idealJ_k(\jac(g,F))$ be the ideal of
	$\pringshort$ generated by the first $k$ columns of $\jac(g,F)$. Then
	assuming that \cref{conj:analysis:froberg} and
	\cref{conj:analysis:U-nonempty} are true, the number of arithmetic
	operations in $\field$ required to compute a grevlex Gr\"obner basis
	for the ideal
	$\critideal=\genby{F}+\maxminorsideal[\nrows+1]{\jac(g,F)}$ is in
	\[
			O\left(\sum_{d=d_0}^{D}\left(\binom{\nvars+d-1}{\nvars-1}-\HF_{\pringshort/\critideal}(d)\right)^{\omega-2}\numrows(d)\binom{\nvars+d-1}{\nvars-1}\right)
	\] 
	with
	\begin{multline*}
		\numrows(d)=\nrows\binom{\nvars+d-d_0-1}{\nvars-1}+\binom{\nvars+d-(\nrows+1)d_0-1}{\nvars-1}\binom{\nvars}{\nrows+1}\\-\left(\sum_{k=1}^{\nvars-\nrows-1}\HF_{\idealJ_k(\jac(g,F))}(d-(\nrows+1)d_0)\binom{\nvars-k-1}{\nrows}\right)
	\end{multline*}
	where the Hilbert function $\HF_{\pring/\critideal}(d)$ is given in
	\cref{prop:analysis:crit-hf}, the Hilbert functions
	$\HF_{\idealJ_k(\jac(g,F))}(d)$ are those given in
	\cref{cor:analysis:columns-hf}, $D=(\nvars+\nrows)d_0+1$, and
	$2\le\omega\le 3$ is a suitable exponent of matrix multiplication.
\end{theorem}
\begin{proof}
	By \cite[Cor.\,2.3]{Spa14}, the largest degree of an element of
	the reduced grevlex Gr\"obner basis of $\critideal$ is
	$D=(\nvars+\nrows)d_0+1$. Therefore, the output of $\CritGB(F,g,D)$
	(see \cref{alg:criticalpointsgb}) is a grevlex Gr\"obner basis of
	$\critideal$. 

	The arithmetic complexity of \cref{alg:criticalpointsgb} is clearly
	bounded by that of its final step. $\SigGB$ only performs arithmetic
	operations when computing row-echelon forms for the Macaulay matrices
	it builds. We can therefore bound the number of arithmetic operations
	performed by $\SigGB$ on any given input by the cost of echelonizing
	the Macaulay matrices it encounters. For a given $d_0\le d\le D$, the
	Macaulay matrix $\macmat_d(\critideal)$ is of rank
	$\HF_{\critideal}(d)$and has $\#\Mon=\binom{\nvars+d-1}{\nvars-1}$. The
	rows of $\macmat_d(\critideal)$ are indexed by a subset of
	\[
		\left(\Mon[d-d_0][\pringshort^{\#F}]\cup\Mon[d-(\nrows+1)d_0][\pringshort^{\#\detsystem{\nrows+1}{\jac(g,F)}}]\right)\smallsetminus H_{d-\nrows d_0}
	,\] 
	which has cardinality precisely $\numrows(d)$ by
	\cref{prop:analysis:num-syz}. By \cite[Sec.\,2.2]{Storjohann2000},
	an $s\times t$ matrix of rank $r$ over $\field$ can be echelonized
	using $O(r^{\omega-2}st)$ operations, so the result follows.
\end{proof}

\subsection{Comparison with Lazard's algorithm}

We conclude with a comparison of the upper bound 
\[
\sum_{d=\nrows d_0}^{d_0(\nrows-1)+(d_0-1)\nvars+1}\numrows(d)
\] 
from \cref{thm:analysis:final-bound} on the total number of rows in all of the
Macaulay matrices built by \cref{alg:maxminorsgb} to the upper bound
\[
\binom{\ncols}{\nrows}\binom{d_0(\nrows-1)+(d_0-1)\nvars+1+\nvars}{\nvars}
.\] 
on the number of rows in the Macaulay matrices built by Lazard's algorithm
obtained in \cite[Theorem\,20]{FaugereSafeySpaenlehauer2013}.

This comparison does not take into account the fact that in \cref{alg:mF5-syz},
Macaulay matrices are computed degree-by-degree, so that reductions to zero in
lower degrees can be used to eliminate reductions to zero in subsequent
degrees.

This comparison also does not take into account the $F_5$ criterion
\cite[Thm.~1]{Faugere2002}, which allows for several more reductions to zero
to be avoided. However, the complexity of $F_5$ has only been analyzed in the
case of a regular sequence, in \cite{BardetFaugereSalvy2015}.

\begin{remark}
	By computing a $\ordpot$-Gr\"obner basis for
	$\Syz(\maxminorssystem{\polymat})$, we can estimate the number of extra
	reductions to zero avoided by \cref{alg:maxminorsgb} thanks to
	\cref{prop:syz-crit} (ii). For $\nrows=3,\ncols=6,\nvars=4,d_0=3$, the
	ratio of the number of rows estimated by taking into account this
	criterion to the number of rows computed by Lazard's algorithm is
	$29.397$, while not taking into account this criterion yields a ratio
	of  $26.786$. Similarly, for $\nrows=3,\ncols=7,\nvars=5,d_0=3$, taking
	into account this criterion gives a ratio of $41.006$, while not taking
	into account this criterion gives a ratio of  $34.946$. This suggests
	that a careful complexity analysis of \cref{alg:criticalpointsgb} might
	provide a theoretical complexity improvement that is better than the
	one suggested by the graphs we give here.
\end{remark}

\subsubsection{$\nrows,\nvars$ fixed, $d_0$ grows}
First, we fix the number of polynomials $\nrows$ and the number of variables
$\nvars$, and allow the degree $d_0$ to grow. We take $\ncols=\nvars+\nrows-1$,
so that the ideal of maximal minors has dimension zero. 
\begin{figure}[ht]
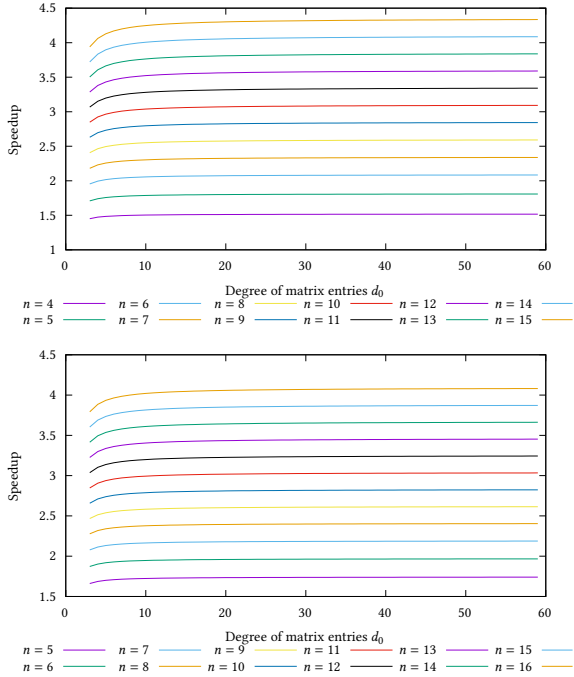

	\centering
	\resizebox{0.9\linewidth}{!}{\input{data/p3_d_grows_lazard_v_us}}
	\resizebox{0.9\linewidth}{!}{\input{data/p4_d_grows_lazard_v_us}}
	\caption{\textnormal{Speedup of \cref{alg:maxminorsgb}. Top: $p=3$; bottom: $p=4$}}
	\label{fig:p3-p4-d-laz-us}
\end{figure}
\Cref{fig:p3-p4-d-laz-us} shows that for a fixed
$\nrows$ and $\nvars$, the theoretical gain which we obtain appears to grow
logarithmically in $d_0$.

Finally, we compare Lazard's algorithm to a (nonexistent) algorithm which would
compute full-rank Macaulay matrices.
\begin{figure}[ht]
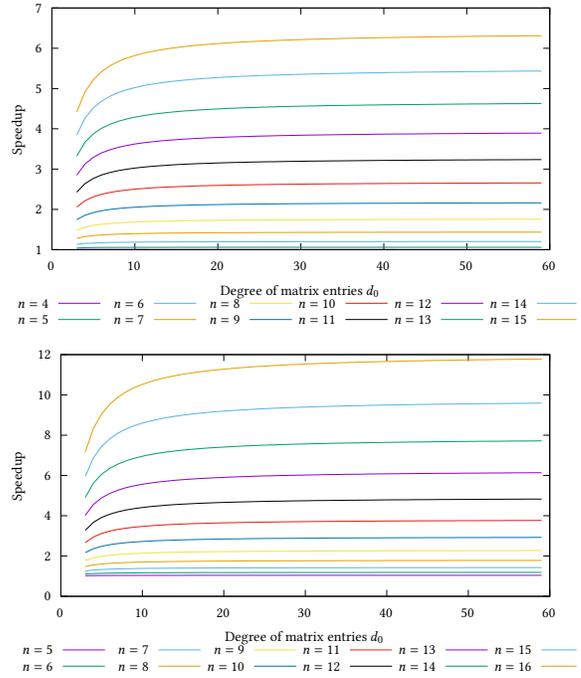

	\centering
	\resizebox{0.9\linewidth}{!}{\input{data/p3_d_grows_lazard_v_en}}
	\resizebox{0.9\linewidth}{!}{\input{data/p4_d_grows_lazard_v_en}}
	\caption{\textnormal{Speedup of an algorithm which computes full-rank
	Macaulay matrices. Top: $p=3$; bottom: \(p=4\).}}
	\label{fig:p3-d-laz-en}
\end{figure}
\cref{fig:p3-p4-d-laz-en} shows that such an algorithm
appears to  also only provide a theoretical gain which grows logarithmically in
$d_0$.

\subsubsection{$\nrows,d_0$ fixed, $\nvars$ grows}
Next, we  fix the number of polynomials $\nrows$ and the degree $d_0$ and allow
the number of variables $\nvars$ to grow. Again, we take
$\ncols=\nvars+\nrows-1$ so that the ideal of maximal minors has dimension
zero.
\begin{figure}[hb]
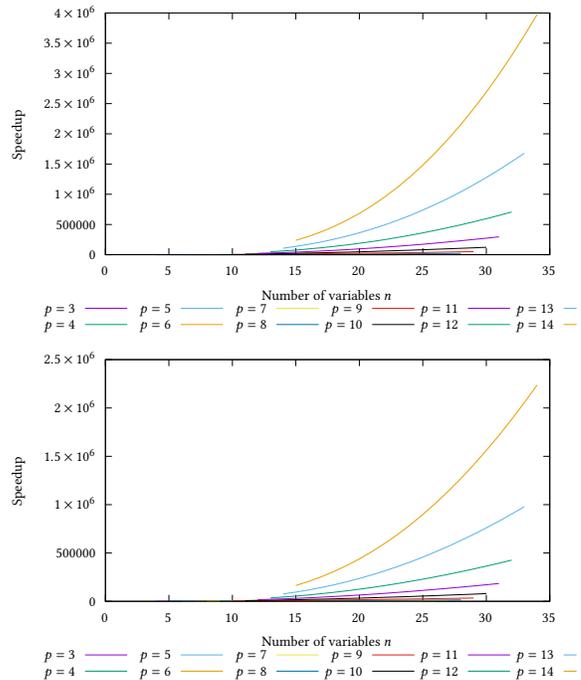

	\centering
	\resizebox{0.9\linewidth}{!}{\input{data/d3_n_grows_lazard_v_us}}
	\resizebox{0.9\linewidth}{!}{\input{data/d4_n_grows_lazard_v_us}}
	\caption{\textnormal{Speedup of \cref{alg:maxminorsgb}. Top: $d_0=3$; bottom: \(d_0=4\).}}
	\label{fig:d3-d4-n-laz-us}
\end{figure}
\Cref{fig:d3-d4-n-laz-us} shows that for a fixed
$\nrows$ and $d_0$, the theoretical gain which we obtain appears to grow
linearly in $\nvars$.

We conclude by again comparing Lazard's algorithm to a (nonexistent) algorithm
which would compute full-rank Macaulay matrices.
\begin{figure}[ht]
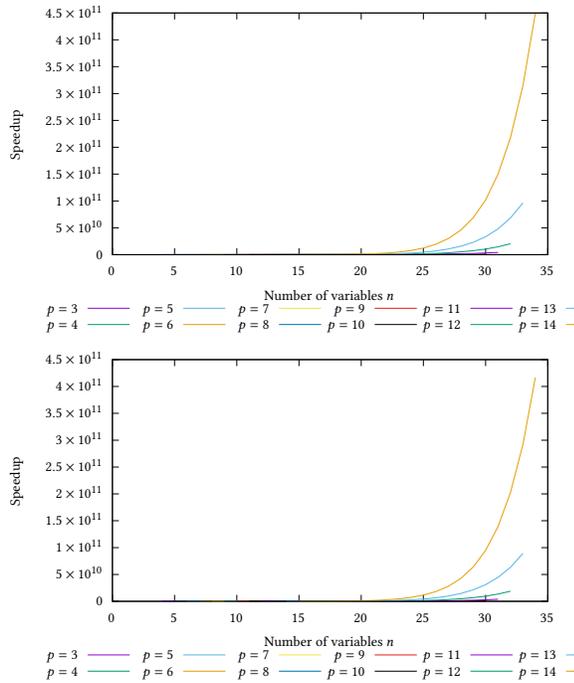

	\centering
	\resizebox{0.9\linewidth}{!}{\input{data/d3_n_grows_lazard_v_en}}
	\resizebox{0.9\linewidth}{!}{\input{data/d4_n_grows_lazard_v_en}}
	\caption{\textnormal{Speedup of an algorithm which computes full-rank
	Macaulay matrices. Top: $d_0=3$; bottom: \(d_0=4\).}}
	\label{fig:d3-d4-n-laz-en}
\end{figure}
\Cref{fig:d3-d4-n-laz-en} shows that such an algorithm
appears to provide a theoretical gain which grows exponentially in $\nvars$,
demonstrating that there is still potentially much to be gained by devising new
criteria which  predict more reductions to zero.

\balance

\end{document}